\documentclass[preprint,12pt,sort&compress]{elsarticle}
\usepackage{graphicx}
\usepackage{amsbsy,amsmath,amsthm,latexsym,amssymb}
\usepackage{color,ulem}
\usepackage{verbatim}  
\newcommand{\bc}{\begin{center}}
\newcommand{\ec}{\end{center}}
\newcommand{\bfr}{\begin{flushright}}
\newcommand{\efr}{\end{flushright}}

\newcommand{\no}{\noindent}
\newcommand{\be}{\begin{enumerate}}
\newcommand{\ee}{\end{enumerate}}
\newcommand{\bi}{\begin{itemize}}
\newcommand{\ei}{\end{itemize}}
\newcommand{\bd}{\begin{description}}
\newcommand{\ed}{\end{description}}

\newcommand{\eeq}{\end{equation}}
\newcommand{\bea}{\begin{eqnarray}}
\newcommand{\eea}{\end{eqnarray}}

\newcommand{\bfi}{\begin{figure}}
\newcommand{\efi}{\end{figure}}
\newcommand{\bay}{\begin{array}{l}}
\newcommand{\eay}{\end{array}}

\newcommand{\cref}[1]{(\ref{#1})}   

\newcommand{\ie}{\textit{i.e.}~}
\newcommand{\ca}{\textit{ca.}~}
\newcommand{\eg}{\textit{e.g.}~}
\newcommand{\via}{\textit{via}~}
\newcommand{\viz}{\textit{viz.}~}
\newcommand{\vs}{\textit{vs.}~}
\newcommand{\eqname}{Eq.~}

\newcommand{\figname}{Fig.~}

\newcommand{\secname}{Section~}

\renewcommand{\thefootnote} 

\begin{document}
\baselineskip 26pt







%
%
%















\pagestyle{empty} \newpage

\pagestyle{plain}

\begin{center}
{\Large {\bf  Nanoparticle simulations of logarithmic creep and microprestress relaxation in concrete and other disordered solids}}
\\[9mm]
{\large {\bf Enrico Masoero$^{\mathrm{a,b}}$, and Giovanni Di Luzio$^{\mathrm{b}}$
}}

\vskip 6mm

{\footnotesize $^{\mathrm{a}}$ School of Engineering, Newcastle University, NE1 7RU, Newcastle upon Tyne, U.K.}\\[1mm]

{\footnotesize $^{\mathrm{b}}$ Department of Civil and Environmental Engineering, Politecnico di Milano, Piazza Leonardo da Vinci 32,
20133 Milan, Italy}\\[1mm]


\end{center}


\bigskip
\no {\bf   Abstract}\,
\vskip 3.5mm
{\small
Ba\v{z}ant's microprestress theory relates the logarithmic basic creep of concrete to power-law relaxation of heterogeneous eigenstresses at the nanoscale. However, the link between material chemistry, nanostructure, and microprestress relaxation, is not understood. To approach this, we use a simple model of harmonically interacting, packed nanoparticles, relaxing with and without external stress. Microprestresses are related to per-particle virial stress heterogeneities. Simulation results show that logarithmic creep and power-law microprestress relaxation emerge from generic deformation kinetics in disordered systems, which can occur in various materials and at various scales. When the interactions are matched to some mechanical properties of C--S--H at the 100 nm scale, the predicted microprestresses have similar magnitude as in Ba\v{z}ant's theory. The ability of our simulations to quantitatively link stress relaxation with nanostructure and chemistry-dependent interactions, provides a pathway to better characterise, extrapolate, and even engineer the creep behaviour of traditional and new concretes. 


%
%
\vskip 30mm \noindent 

\noindent\textcopyright ~ 2020. This manuscript version is made available under the CC-BY-NC-ND 4.0 license 

https://creativecommons.org/licenses/by-nc-nd/4.0/

\noindent Published article at https://doi.org/10.1016/j.cemconres.2020.106181 
}

\noindent
\bigskip


\section{Introduction}\label{secIntro}

The long-term basic creep of concrete implies a logarithmic increase of deformations during the service life of structures \cite{bavzant2018creep}. This has sometimes a beneficial effect, such as to accommodate imposed strain, \eg due to humidity cycles, thermal gradients, or ground settlements. However, in most cases, creep creates problems such as loss of cable tension in prestressed concrete or even structural collapse \cite{bavzant2012excessive}. Current objectives are to design concrete mixes with pre-established creep behaviours and to extrapolate long-term creep performance from relatively short-term experiments, to help quality control, monitoring, and management of infrastructure \cite{bavzant2018creep}. In both cases, the underlying scientific challenge is to understand the microscopic mechanisms that govern the logarithmic creep of concrete, and how these are determined by the chemical composition and microstructure of the material.

Bazant et al.~\cite{bazant1997microprestress} proposed the microprestress relaxation theory as a mechanistic foundation to model the logarithmic creep of concrete. The theory starts with two assumptions: (1) creep strain originates from shear slips in water-rich layers at the nanometre scale, also called creep sites, that are randomly oriented within calcium-silicate-hydrate (C--S--H); (2) C--S--H features a spatially heterogeneous field of self-equilibrated (eigen)stress, originating when the material forms from chemical and physical processes such as disjoining and cristallisation pressures. The eigenstresses put some creep sites under local tensile stress and others under compression. Sites under tension are more likely to slip under shear and, when this happens, the local rearrangement triggers a broader relaxation of eigenstress across the material. In turn, this relaxation reduces the local tensile stress at other sites, decreasing the rate of subsequent slips, and thus the creep rate. All these assumptions align with traditional \cite{feldman1972mechanism} and modern  understanding of shear slips in C--S--H at the molecular scale \cite{manzano2013shear,vandamme2015creep,morshedifard2018nanoscale}, and led to models that can fit the experimental results. However, the mechanism of microprestress relaxation has neither been directly observed nor simulated to date.

Recent studies have started to investigate the  mechanisms of logarithmic creep in C--S--H at the nanoscale. Nanoindentation experiments have shown that, at length scales below the micrometre, logarithmic creep emerges already over short time scales \cite{vandamme2009nanogranular}. The result has been interpreted as viscous compaction of a nanogranular solid, using the framework of free volume theory \cite{cohen1959molecular}. Molecular and nanoparticle simulations have predicted logarithmic creep to emerge from plastic deformations in disordered structures \cite{masoero2013kinetic,bauchy2015creep,bauchy2017topological,morshedifard2018nanoscale,liu2019long}. However, stress heterogeneities have not been analysed in those simulations, nor they can be accessed in nanoindentation experiments. These recent insights are still to be reconciled with the microprestress relaxation theory.

Here we present a simple model to simulate microprestress relaxation in disordered porous materials. The solid is discretised using particles that interact \via a spherical harmonic potential, here parametrised to reproduce the elastic properties of C--S--H at the 100 nm scale. Microprestresses are expressed as functions of per-particle virial stress heterogeneities. Simulation results predict logarithmic creep and power-law relaxation of microprestress, confirming Ba\v{z}ant et al.'s theory and linking it quantitatively to the nanostructure and, \via the interaction potential, to the chemical composition of the material.

\section{Methodology}\label{secMethod}

\subsection{Elements of microprestress relaxation theory}

The microprestress theory is now summarised. Let us consider a disordered material featuring a heterogeneous field of self equilibrated eigenstress, $\Sigma(\textbf{r})$, where $\textbf{r}$ is the position vector. {\color{black}Ba\v zant et al.~\cite{bazant1997microprestress} identified the microprestress with the} average tensile eigenstress, $S = \frac{1}{V^+}\int_{V^+} \Sigma(\textbf{r})dV$, where $V^+$ is the portion of material's volume under tensile eigenstress. {\color{black} The rationale to disregard local compressive stresses is that they increase the activation energy for shear slips \cite{vandamme2015creep}, thus reducing exponentially the probability that such slips could originate at sites under compression.} The creep rate is $\dot\varepsilon = \tau/\eta(S)$ where $\tau$ is the external stress driving creep and $\eta(S)$ is the viscosity of the material. Ba\v zant et al.~proposed an expression for $\eta(S)$ based on self-similarity and recently confirmed by simulation \cite{vandamme2015creep}: $\frac{1}{\eta(S)} = cpS^{p-1}$, where c and p are constants. A last equation governs the temporal relaxation of $S$: 
\begin{equation}\label{eqMPeq}
\frac{\dot{S}}{C_s} + \frac{S}{\eta(S)} = 0
\end{equation}
$C_s$ is an elastic constant. \eqname(\ref{eqMPeq}) assumes randomly oriented planes whose slip rate (second term) equals the rate at which stress relaxation causes deformation at other planes \via the elastic medium (first term). The solution of \eqname(\ref{eqMPeq}) is a power law:
\begin{equation}\label{eqSpow}
S = S_0 \left(\frac{t}{t_0}\right)^{-\alpha}
\end{equation}
with $\alpha = \frac{1}{p-1}$ and $S_0$ being $S$ at the arbitrary time $t_0$. Ba\v zant et al.~originally proposed $p = 2$, thus $\alpha = 1$, but any $\alpha > 0$ would lead to logarithmic creep when substituting $S(t)$ into $\eta(S)$ and then into $\dot\varepsilon$. 

Hereafter, to compute strain rate and eigenstress relaxation, we develop a simple model of dense amorphous material, impose a field of eigenstress, and then perform accelerated creep simulations under constant shear stress.

\subsection{Particle-based model description}\label{SecModDesc}

Our simple model is a binary mixture with few large spherical particles in a matrix of smaller ones. The particles interact \textit{via} a pairwise size-dependent harmonic potential:
\begin{equation}\label{eqUij}
U_{ij} = \frac{1}{2}k\left(r_{ij}-{D_{ij}}\right)^2 - U_{0,ij}
\end{equation}
$r_{ij}$ is the distance between particles $i$ and $j$, whose equilibrium distance is their average diameter ${D_{ij}}$. A cutoff is applied, such that $U_{ij} = 0$ when $r_{ij} \ge r_u$. The term $U_{0,ij}$ is the separation energy from $r_{ij} = {D_{ij}}$ to $r_{ij} > r_u$. The potential therefore has three mechanical parameters: $k$, $r_u$, and $U_{0,ij}$. Harmonic potentials are widely used in nanoscale simulations as they provide the simplest model for inter-atomic bonds as well as inter-particle cohesion in C-S-H and other materials \cite{mishra2017cemff,harrison2018review,poschel2005computational}.

Later we will compare results with the microprestress theory of concrete, thus we parametrise the model to capture some mechanical properties of the C--S--H phase at the 100 nm scale. Particle diameters $D$ are set to 5 and 7 nm \cite{allen2007composition,Masoero2012}. Assuming perfect cohesion between particles, we set $k = \frac{EA_{ij}}{{D_{ij}}}$, where $E = 63.6$ GPa is the elastic modulus of C--S--H at the molecular scale, and $A_{ij} = \frac{\pi}{4}{D_{ij}}^2$ is the contact area between two particles \cite{masoero2014modelling}. The cutoff is set to $r_u = \varepsilon_u {D_{ij}}$, where $\varepsilon_u = 0.03$ is a reasonable strain at tensile failure for C--S--H at the molecular scale, {\color{black} as indicated by molecular simulations \cite{murray2010molecular,hou2015uniaxial,xin2017temperature} and consistent with experimentally measured values of nanoindentation modulus and hardness \cite{Constantinides2007,Pellenq_al_PNAS_2009}}. $U_{0,ij}$ is set to $2\gamma A_{ij}$, where $\gamma = 87.6$ mJ m$\mathrm{^{-2}}$ is the interfacial energy between C--S--H and its surrounding solution in concrete \cite{bullard2015time,shvab2017precipitation}.

We construct six configurations starting from two statistically equivalent baseline structures, $A$ and $B$, featuring \ca $10,000$ particles each (see \ref{AppA} for more details). The baseline structures are amorphous, monodisperse ($D=5$ nm for all particles), and dense (packing densities $\eta_A = 0.63$ and $\eta_B = 0.62$, nearing the 0.64 limit of random close packing for monodisperse hard spheres). Their average XYZ axial stresses, computed with the virial method, are set to zero by combining changes of XYZ box sizes with energy minimisation and random agitation (see \ref{AppA} for more details). The agitation also reduces the stress heterogeneities, towards minimum albeit nonzero values. Subsequently, a new and intense field of eigenstress is introduced by inflating a fraction $\delta$ of particles, whose diameters are increased to 7 nm. {\color{black} Particles inflation mimics a generic set of local expansive processes, which in Ba\v zant et al.'s theory are the source of the microprestress: \eg crystallization pressure from hindered precipitation of solids, or disjoining pressure from hindered expansion of fluid in the nanopores$^1$.}
\footnote{\color{black} $^1$ The pore solution is not modelled explicitly, but the interaction potential between particles is typically considered to represent the mechanics of a water-rich interlayer space in the C--S--H \cite{masoero2014modelling,ioannidou2016mesoscale,bonnaud2016interaction,masoumi2020nanolayered}. The effect of fluid expansions, causing disjoining pressure if hindered, can therefore be modelled as an increase in equilibrium distance between two interacting particles, \ie an inflation of the diameter $D_{ij}$ in \eqname\ref{eqUij} as we did here.}
We create structures with $\delta = 5\%$, $10\%$, and $20\%$ from both baselines A and B, obtaining a total of six configurations. Particle inflation generates a large average pressure, which is zeroed again \via energy minimisations and changes of XYZ box sizes. Differently from the baseline structures, now we want to preserve an intense field of eigenstress to be relaxed later, therefore now we do not apply random agitation after inflation.

We measure the eigenstress per particle $\Sigma$ as the hydrostatic stress $\frac{1}{3}\mathrm{Tr}{\bf \Sigma}$, where ${\bf \Sigma}$ is the 3$\times$3 virial stress tensor per particle \cite{thompson2009general} obtained taking $V/N$ as volume per particle ($V$ is the volume of the simulation box, $N$ the number of particles). With this definition, $\Sigma$ is a measure of local compression or tension. The average tensile eigenstress can be computed as $s = \frac{1}{N^+}\sum_{N^+} \Sigma_i^+$, where $N^+$ is the number of particles with $\Sigma_i^+ > 0$. Here $s$ is not the same as the microprestress $S$ in Ba\v zant et al.~\cite{bazant1997microprestress}; we will see later that the two are closely related, but not identical.

\begin{figure}[h]
\includegraphics[width=0.8\columnwidth]{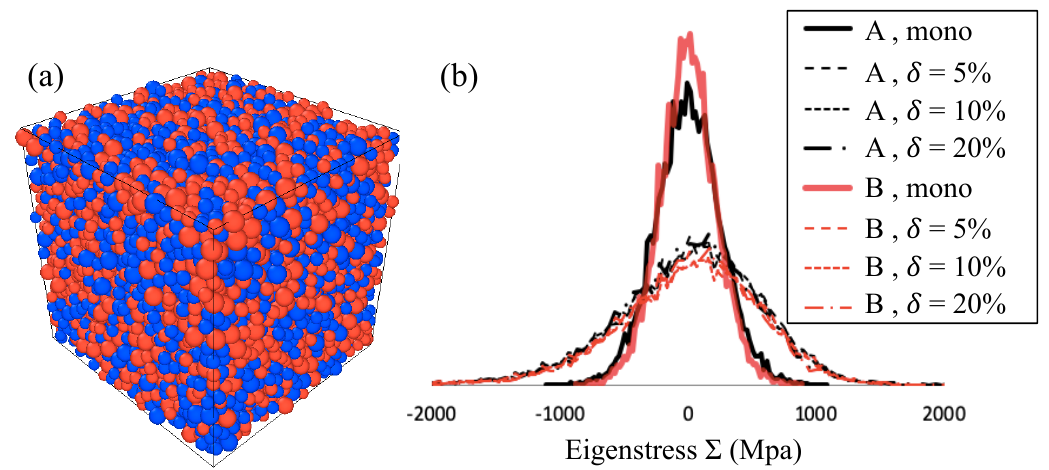}
\caption{(a) OVITO \cite{stukowski2009visualization} snapshot of structure B with $\delta = 20\%$ of inflated particles: blue (dark) particles have $\Sigma < 0$, \ie are under local compression; red (light) particles are under tension, $\Sigma > 0$. (b) Eigenstress distributions at zero average axial stresses for configurations A and B, both before (monodisperse) and after inflating different fractions $\delta$ of particles. The average positive $\Sigma$ before inflation are $s_A=101$ MPa and $s_B=85$ MPa. After inflation, the distributions for all $\delta$'s almost coincide, with $s_i$ all between 206 and 219 MPa. The subscript in $s_i$ indicates that these are the initial eigenstresses for subsequent simulations of creep and relaxation.}\label{Fig1}
\end{figure}

\figname\ref{Fig1}.b shows the distributions of eigenstress in the six configurations, after particle inflation and stress minimisation but before simulating creep and eigenstress relaxation. The distributions are compared to those in the two monodisperse baseline structures, just before particle inflation. \figname\ref{Fig1}.b shows that particle inflation intensifies the eigenstress field, as indicated by the widening of the distribution tails. The distributions after inflation are very similar for all $\delta$'s. This suggests that increasing diameters from 5 to 7 nm causes local yielding, which caps the local eigenstress to a maximum. Smaller values of $s$ could be obtained by inflating fewer particles ($\delta < 5\%$) or by increasing their diameters less. 

\subsection{Simulating creep and eigenstress relaxation}\label{SecMethCreep}

The six numerical model structures, under intense eigenstresses, are first tested for creep. We use an oscillatory shear protocol analogous to previous simulations of the logarithmic creep of C--S--H \cite{bauchy2015creep,bauchy2017topological,liu2019long}. A constant shear stress $\tau_{xy} = 40$ MPa is imposed. {\color{black} Then the simulation box is frozen and a cycle of shear strain,} with $\Delta \gamma_{xy} = \pm 0.03$, excites the system. {\color{black} When $\Delta \gamma_{xy}$  is applied, the shear stress changes by a corresponding $\Delta \tau_{xy}$. If the system's response to strain was linear elastic, $\Delta \tau_{xy}$ would equal $G\Delta \gamma_{xy} \approx 180$ MPa, where $G \approx 6$ GPa is the shear modulus (see \ref{AppB}). Actually, nonlinearities and even irreversible rearrangements are expected during the strain cycle, so 180 MPa is an upper bound for $\Delta \tau_{xy}$}.
The values of $\tau_{xy}$ and $\Delta\gamma_{xy}$ are decided based on quasi-static shear tests (see details in \ref{AppB}), following two principles: (i) $\tau_{xy}$ must be significantly smaller than the yield stress, and (ii) the {\color{black} upper bound} shear stress during a cycle, $\tau_{xy} + G\Delta \gamma_{xy}$, must be close but still smaller than the yield stress {\color{black} (a too small $\Delta \gamma_{xy}$ would lead to excessively rare rearrangements, making the activation ineffective over the timescale of a simulation; a too large $\Delta \gamma_{xy}$ would trigger system-spanning rearrangements, \eg shear bands, that are typical of yielding but that are unrealistic deformation mechanisms for creep)}. After each strain cycle the interaction energy of the system is minimized, while also adjusting the XYZ box dimensions and the XY angle, until average axial stresses are null and $\tau_{xy}$ is back to 40 MPa. The irreversible shear strain, which increases with the number of cycles, is the creep strain $\gamma_{xy}$.

In the microprestress theory \cite{bazant1997microprestress} relaxation occurs even when there is no external stress applied, and this explains why the basic creep compliance of concrete decreases as the material ages. To simulate relaxation without external load ($\tau_{xy} = 0$), we compare two relaxation protocols: (i) the same oscillatory strain protocol as for the creep tests above, with $\tau_{xy} = 0$ and $\Delta \gamma_{xy} = \pm 0.04$; (ii) cycles of 500-1,000 steps of accelerated molecular dynamics (AMD), with a Nose-Hoover thermostat applying random velocities consistent with average kinetic energy per particle $e_k = 0.15~U_0$ (here $U_0$ is the separation energy for particles with $D=5$ nm). The AMD cycles are carried out at constant volume, but after each cycle the XYZ box dimensions are changed to restore zero average axial stresses {(\color{black} alternatively, we could have let the box dimensions and shape change during the AMD cycles, using a barostat to keep the stresses constant: the results would have been different but statistically equivalent)}. The values of $\Delta \gamma_{xy}$ and $e_k$ have been chosen to maximise eigenstresses relaxation without causing system-wide damage (see details in \ref{AppC}).

\section{Results and Discussion}\label{secRes}

\subsection{Creep simulations and corresponding microprestress relaxation}

\begin{figure}[h]
\includegraphics[width=0.7\columnwidth]{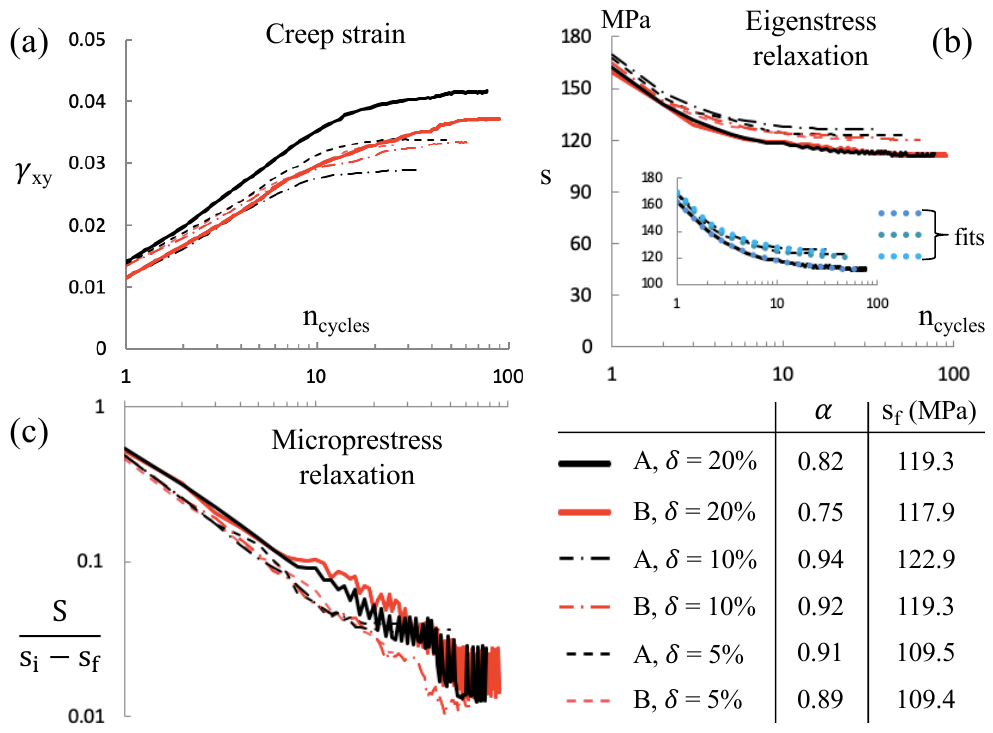}
\caption{Results of creep simulations under shear: (a) strain, (b) average tensile eigenstress $s$, and (c) microprestress $S = s-s_f$ normalised by $s_i-s_f$, all as functions of the number of shear strain oscillations $n$. $s_i$ is the initial value of $s$ before the strain oscillations, as discussed in the caption of \figname\ref{Fig1}. The logarithmic regime in (a) corresponds to a power law regime of microprestress relaxation in (c). The table shows key parameters to fit the curves in (b) using \eqname\ref{eqsfit}. The inset in (b) shows the good quality of the fits for configurations of type A; similar quality holds also for types B.}\label{Fig2} 
\end{figure}

\figname\ref{Fig2} shows the results of creep simulations. In \figname\ref{Fig2}.a, all configurations start with an initial logarithmic regime of strain \vs number of perturbative strain cycles, $n$. Following \cite{bauchy2017topological}, we use $n$ as a proxy for time by considering that strain perturbations mimic rare thermal fluctuations. The curves in \figname\ref{Fig2}.a are all in the same range, with no trend as a function of $\delta$. As discussed in \secname\ref{SecModDesc} and shown in \figname\ref{Fig1}.b, local plastic deformations upon inflation cap the eigenstresses to similar values in all the considered structures, therefore similar creep behaviours can be expected if the microprestress theory is valid.

\figname\ref{Fig2}.b shows that creep strain is indeed accompanied by eigenstress $s$ relaxation. Also here, all configurations behave similarly. Eigenstress relaxation saturates towards a minimum value for large $n$, which concurs with the end of the logarithmic creep regime in \figname\ref{Fig2}.a, when $\gamma_{xy}$ stabilises towards a maximum. A more extended logarithmic regime could be obtained using more advanced protocols than the simple oscillatory one used here, \eg stress marching \cite{morshedifard2018nanoscale}. Indeed, previous creep simulations on a similar model featuring more size polydispersity, predicted logarithmic creep over $n=10^6$ cycles \cite{liu2019long}. Here, however, it is useful to study the systems as they leave the logarithmic regime and see that this is accompanied by saturation of eigenstresses. {\color{black} If such correspondence between regimes had not emerged, \viz if logarithmic creep continued despite eigenstress saturation or if eigenstress relaxation proceeded as a power law despite a change in creep regime, then the microprestress relaxation mechanism could not have explained our simulation results.}

Differently from the microprestress $S$ in Ba\v zant et al.~\cite{bazant1997microprestress}, the eigenstress $s$ in \figname\ref{Fig2}.b is not a power law of $n$. Indeed, $s$ does not tend to zero as $n\rightarrow\infty$, but rather to a finite value $s_f>0$. The impossibility to achieve $s=0$ is to be expected in dense, amorphous, frustrated systems of interacting particles. {\color{black} More in detail, the interaction potential in \eqname\ref{eqUij} generates strong repulsive forces when $r_{ij}<D_{ij}$, turning the particles into excluded volume for rearrangements, and neglecting viscous process at the molecular scale, inside the particles, which could sustain further relaxation below $s_f$. The original microprestress theory, by contrast, stemmed from a continuum-based description, where the resolution of displacements was the infinitesimal volume element. In that context, it was consistent to assume $s_f = 0$. The reality for C--S--H probably falls in-between, as strong repulsions causing excluded volumes are expected at the atomic scale (\ca 0.1 nm), which is between our particles' scale (5-7 nm) and the infinitesimal volume. Therefore, C--S--H should feature a finite $s_f$ although smaller than our simulations predict.}

To recover a measure of microprestress that tends to zero, we simply define {\color{black} it as} $S = s-s_f$. {\color{black} In this way, we identify the microprestress $S$ only with the part of the eigenstress $s$ that can relax, causing the viscosity to change and the logarithmic creep to develop. Ours is therefore a generalisation of the original definition from Ba\v zant et al., which is recovered when $s_f=0$}. Substituting our $S$ into \eqname\ref{eqSpow} and solving for $s$, one obtains an equation that should fit $s(n)$ in \figname\ref{Fig2}.b, provided that $S$ in our simulations is indeed a power law of $n$:
\begin{equation}\label{eqsfit}
s = S_0 \left(\frac{n}{n_0}\right)^{-\alpha} + s_f
\end{equation}
$s_f$, $\alpha$, and $S_0$, are three parameters that we compute by least square fitting of the curves in \figname\ref{Fig2}.b. 

The excellent quality of the fits, in the inset of \figname\ref{Fig2}.b, confirms the assumption that our simulated microprestress $S$ relaxes as a power of $n$, as shown in \figname\ref{Fig2}.c. The fitted values of $\alpha$ and $s_f$ are tabulated in \figname\ref{Fig2}; $S_0$ is not shown because it is not informative, as it depends on an arbitrary $n_0$. The values of $\alpha$ and $s_f$ are similar for all configurations A and B, with no trend as a function of $\delta$. 
The simulated exponents $\alpha$ are close to $\alpha = 1$, which Ba\v zant et al.~\cite{bazant1997microprestress} proposed by fitting creep experiments on concrete. Another quantitative agreement comes from the total relaxed micropresesses, $s_i-s_f$ ($s_i$ is discussed in the caption of \figname\ref{Fig1}), which amounts to \ca 95 MPa for all our six configurations. This is smaller but not excessiely far from the 150-200 MPa that Ba\v zant et al.~\cite{bazant1997microprestress} suggested as an upper-bound for C--S--H. Overall, \figname\ref{Fig2} shows that the adopted nanoscale model of amorphous material displays logarithmic creep and power law relaxation of excess stress heterogeneities, as inferred in the microprestress theory.

\subsection{Microprestress relaxation without external load (ageing)}

\begin{figure}[h]
\includegraphics[width=0.9\columnwidth]{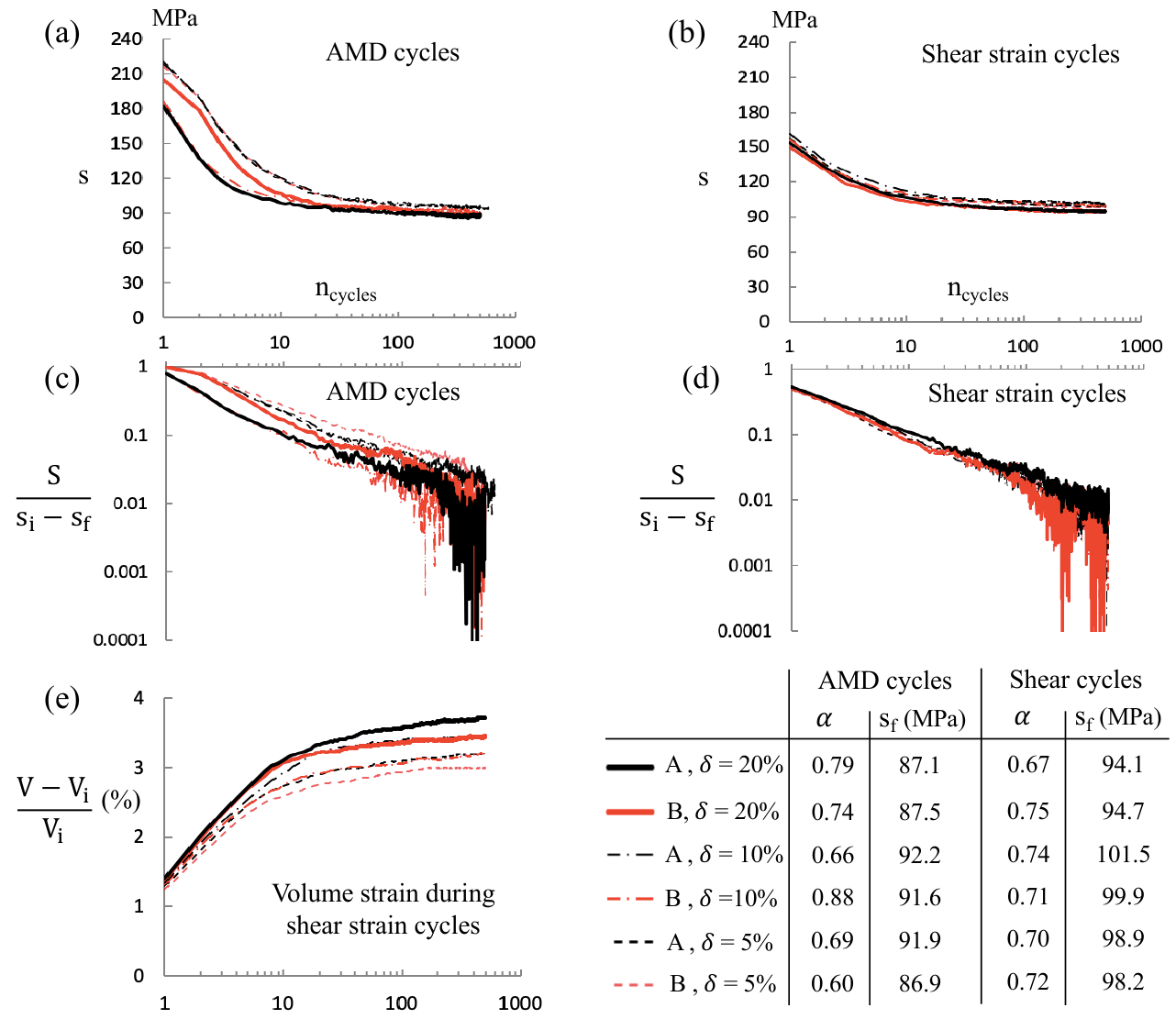}
\caption{Relaxation of eigenstress $s$ and microprestress $S=s-s_f$ when no external stress is applied. Perturbations induced by (a,c) accelerated molecular dynamics, AMD, and (b,d) shear strain oscillations. The table shows key parameters to fit the curves in (a,b) using \eqname\ref{eqsfit}. (e) Volume expansion accompanying relaxation by shear strain oscillations. $V_i$ is the initial volume of a configuration at $n_{cycles}=0$.}\label{Fig3} 
\end{figure}

\figname\ref{Fig3} presents the results of eigenstress relaxation in the numerical simulations when the externally applied stress is zero, $\tau_{xy} = 0$. As detailed in \secname\ref{SecMethCreep}, we employed two protocols to carry out such simulations: accelerated molecular dynamics (AMD) and cyclic shear. The results in \figname\ref{Fig3} show that both protocols induce similar eigenstress relaxation. The results are qualitatively similar to those in \figname\ref{Fig2} and also the fitted parameters are similar, except for some small but meaningful differences. In particular, the asymptotic minimum $s_f$ in \figname\ref{Fig3} are \ca 20\% smaller than those in \figname\ref{Fig2}, which is expected because the simulations in \figname\ref{Fig3} have been calibrated to maximise eigenstress relaxation, as explained in \secname\ref{SecMethCreep}. The smaller $s_f$ therefore are explained by the larger strain perturbation $\Delta \gamma_{xy}$ and by the fact that the AMD protocol in \figname\ref{Fig3}.a aims to the same state of maximum relaxation. Smaller $s_f$ in \figname\ref{Fig3} entail more relaxation of microprestress and indeed $s_i-s_f$ is now \ca 115 MPa, which is even closer than before to the 150-200 MPa upper bound theorised in Ref.~\cite{bazant1997microprestress}.

The power law exponents $\alpha$ are also \ca 17\% smaller in \figname\ref{Fig3} compared to \figname\ref{Fig2}. This indicates differences in the deformation mechanisms during relaxation. A possible explanation is that stronger perturbations in \figname\ref{Fig3} might cause more particles to be involved in the rearrangements that underpin the accumulation of irreversible deformations $\gamma_{xy}$. Another possibility is that the external stress $\tau_{xy}$ has an effect on $\alpha$. Ba\v zant et al.~argued that the external load should only negligibly impact microprestress relaxation, but only because creep experiments on concrete typically use loads of 10 MPa or less, which is small compared to relaxations over $100$ MPa. In our creep tests, instead, $\tau_{xy} = 40$ MPa is comparable to the eigenstress, and this might alter the deformation mechanisms. These first results create scope for future research into the details of the deformation mechanisms and their dependence on external stress and relaxation protocols.

\subsection{Volumetric changes accompanying microprestress relaxation}

The free volume theory explains logarithmic creep of disordered porous materials as the result of compaction, which progressively reduces the probability of further compaction to occur, thus causing the strain rate to decrease \cite{cohen1959molecular,kassner2015creep}. This theory has been invoked to interpret experimental measurements of logarithmic creep from short-term nanoindentation tests \cite{vandamme2009nanogranular}. Interestingly, however, \figname\ref{Fig3}.e shows an opposite trend: the volume increases during eigenstress relaxation, and a similar expansion also emerged during the creep tests in \figname\ref{Fig2}. This result might seem counter-intuitive, as relaxation of tensile eigenstress, taken by itself, should indeed cause compaction. However, the system also features compressive eigenstresses that are also relaxing to keep constant the average axial stresses. Following the same logics as above, one should therefore expect no volume changes at all. However, volume changes are not just reflections of eigenstress relaxation in a linear elastic medium, but rather the result of large local deformations caused by particle rearrangements. Such mechanisms are non-linear and their non-trivial impact on the overall volume depends on the details of the interaction potentials.


\subsection{Implications at other length scales and for other materials}

Our results show that microprestress relaxation and logarithmic creep originate from collective shear slips in C--S--H at the 100 nm scale.  However, one could apply the same model and obtain similar results using different particle diameters and interaction parameters. This means that similar mechanisms can emerge also in other disordered materials and at other length-scales.

In the recent literature, logarithmic creep of C--S--H has been predicted both by molecular simulations of interlayer water \cite{bauchy2015creep,bauchy2017topological,morshedifard2018nanoscale} (in spaces where water features glassy structure and kinetics \cite{manzano2012confined}) and by nanoparticle simulations like in the present work \cite{masoero2013kinetic,liu2019long}. This raises the question of whether the decay of strain rate during creep stems from processes at the sub-nano scale or at the nano-to-micro mesoscale. Our results supports the latter interpretation, but the same model could be a first approximation for glassy interlayer water too, as long as particle diameters and interaction parameters are adjusted accordingly (\eg the model used in \cite{vandamme2015creep}). Therefore, analogous deformation mechanisms, eigenstress relaxation, and logarithmic creep, might also emerge at the molecular scale. The methodology presented here provides an approach to test this possibility also using more detailed models for molecular simulations of confined water in C--S--H.

Models similar to ours are commonly used to simulate deformations in glasses, metallic alloys, and wet granular matter, all of which display logarithmic creep in certain conditions \cite{huang2009indentation,darnige2011creep,siebenburger2012creep}. The basic deformation mechanisms in these materials are localised shear rearrangements of multiple nano-units, which can be atoms, molecules, or particles, depending on the system. In the physics community, these rearrangements are known as Shear Transformation Zones (STZs) \citep{falk1998dynamics}. STZs were first proposed by Argon \citep{argon1979plastic} and now there are specific equations of motions describing their formation, disappearance, and activation \cite{bouchbinder2009nonequilibrium}. What is not understood is how the dynamics of STZs can produce the $\dot\varepsilon \sim t^{-1}$ scaling of logarithmic creep. Recent Kinetic Monte Carlo simulations on a simple lattice model have shown that this scaling can emerge from STZs interacting mechanically with each other \via stress redistributions at the mesoscale \cite{bouttes2013creep}. This is analogous to the microprestress relaxation mechanism simulated here. STZs therefore can be regarded as a generalisation of the creep sites in C--S--H and thus the microprestress theory might actually explain the logarithmic creep of a wider class of disordered materials.

\section{Conclusion}\label{secConc}

The numerical simulations presented here have identified nanoscale rearrangements under shear as a mechanism for power-law relaxation of eigenstress heterogeneities and logarithmic creep, as proposed by Ba\v zant et al.~in the microprestress relaxation theory \cite{bazant1997microprestress,bavzant2018creep}. Other hypotheses of the original theory have been confirmed as well, in particular: (i) relaxation of microprestresses occurring also without externally applied stress, as an ageing mechanism, (ii) the order of magnitude of microprestress relaxation, \ca 100 MPa.
For the interaction potential used here, the simulations have indicated that microprestress relaxation is accompanied by volume expansion. This means that logarithmic creep originates from large local deformations, causing non-linear relationships between stress and volume that are not simply associated with compaction. Due to the generic nature of our model, the results in this manuscript can be translated to other material systems also at other scales, as discussed for interlayer C--S--H water at the molecular scale and the deformations of metallic alloys and glasses. In conclusion, this manuscript has presented a pathway to investigate the relationship between creep of disordered materials, relaxation of stress heterogeneities, microstructure, and chemical composition (reflected by the interaction potentials). This provides new opportunities for understanding, extrapolating, and even designing the creep behaviour of ordinary and new concretes.




\bibliographystyle{unsrt}
\bibliography{./bibliocement}


\clearpage

\appendix
\section{Preparation of baseline structures and particle inflation} \label{AppA}

Monodisperse baseline structures are prepared in three steps: (i) a space filling algorithm, (ii) changes of XYZ box sizes to minimise the average axial stresses, and (iii) random agitation to relax stress heterogeneities while also adapting the box to preserve zero axial stress.

The space filling algorithm is a loop whose generic step starts with the insertion of 2,500 trial particles at random locations (our simulation box is initially a cube with edge of 20$D$, where $D$ is the particle diameter). Only trial particles that do not excessively overlap with previously existing particles are accepted and relabelled as ``existing''; the others are rejected and deleted. Our definition of ``excessive overlap'' is when the distance from any existing particle is smaller that $\xi D$, where $\xi$ is a user-decided parameter. $\xi=1$ would mean than only strictly non-overlapping trial particles are accepted. By contrast, smaller $\xi$ increase the rate of particle acceptance, but also build up high mechanical stress. In this manuscript, we used $\xi = 0.7$ for baseline structure A, and $\xi=0.75$ for B. After having converted the trial particles that do not overlap excessively into existing ones, and deleted the rest, the interaction energy of the system is minimised at constant volume, using the conjugate gradient algorithm in LAMMPS. The loop of trial particles insertion, acceptance, and minimisation, is repeated 1,000 times, which is sufficient to obtain densely packed structures with the $\xi$ we adopted, \viz structures with packing density $\eta>0.64$, .

The second step in the preparation procedure is to set the average axial stresses to zero in all three directions. If $\xi<1$ is used, the structure produced by the space filling algorithm is under compressive average axial stresses $\sigma$ in all directions. To relax these $\sigma$, we run a loop where each step consists of one adjustment of all three box lengths, in X, Y, and Z directions. The adjustment of each box edge is determined as $L_{j,new} = L_{j,old}\left( 1-\frac{\sigma_j}{K}\right)$, where $L_j$ is the box length in direction $j=X,Y,Z$, and $K$ is a user-decided constant. The equation implies that a structure under tension in direction $j$, \viz $\sigma_j>0$, will have its $L_j$ reduced by a strain that is proportional to $\sigma_j$. Vice versa, if $\sigma_j<0$, then $L_j$ will increase. The value of $K$ controls the quality and speed of the convergence to zero stress. In our simulations, we used $K = 10^6$ MPa. Each box adjustment is accompanied by affine displacement of all particles in the box, and is immediately followed by minimisation of the interaction energy. After one adjustment and minimisation is performed in all three directions, the new $\sigma_j$ are checked and the loop is broken if $\sigma_x^2+\sigma_y^2+\sigma_z^2<\sigma_{tol}^2$. Here we used tolerance $\sigma_{tol}=10$ MPa, which is small compared to the hundreds of MPa of material strength (see next appendices). 
  
The third step in preparing the baseline structures is to relax the heterogeneous local stresses while keeping zero average axial stresses. To this end we repeat for 1,000 times a loop consisting of two parts: 10,000 steps of accelerated molecular dynamics (AMD) with random agitation at constant volume followed by XYZ box length changes resetting $\sigma$ to zero. The random agitation in the AMD part is applied using the Nose-Hoover thermostat in LAMMPS, targeting random velocities consistent with average kinetic energy per particle $e_k = 0.3~U_0$ (see \secname\ref{SecMethCreep} for the definition of $U_0$ for particles with $D=5$ nm). The XYZ box length changes follow exactly the same protocol as in the previous paragraph.

All the described box changes alter the packing density $\eta$ that the structure displayed just after space filling. Here we targeted a final $\eta$ of 0.64, which is the random close packing limit for monodisperse hard spheres. This target drove a trial-and-error adjustment of parameters which eventually led to $\xi$ = 0.7-0.75, insertion of 2,500 trial particles at each of 1,000 space filling steps, and the above-mentioned intensity $ek$ of random agitation. A systematic exploration of the relationship between preparation parameters and $\eta$ would be interesting, but this was not our priority here.

With monodisperse structures, spherical interactions, and random agitation, there is the risk of inducing local crystallisation. However, this did not occur here, since we checked that the radial pair distribution function featured the split second peak typical of amorphous monodisperse systems, and a common neighbour analysis using OVITO showed that only 0.5\% of all particles had crystalline local environments of FCC or HCP types.  

After completing space filling and eigenstress relaxation in the monodisperse baseline structures, microprestresses are introduced by inflating a fraction $\delta$ of particles, whose diameter is increased from 5 to 7 nm. Particle inflation is carried out at constant volume, thus it generates high compressive stress in all directions. A minimisation at constant volume is immediately performed, which already reduces a bit the average axial stresses, due to local particle rearrangements. However, to actually recover zero axial stresses in XYZ, we must use again the protocol of box length changes described previously in this section. Random agitation is not applied at this stage to preserve an initial field of eigenstress.

\section{Shear stress-strain curves to decide $\Delta\gamma$ and $\tau$ in the simulations of creep} \label{AppB}

\begin{figure}[h]
\includegraphics[width=0.7\columnwidth]{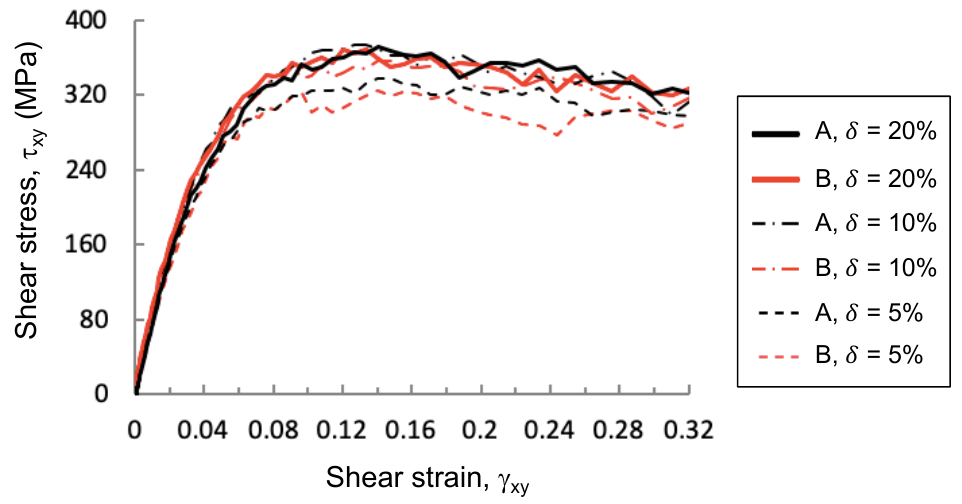}
\caption{Shear stress-strain curves for all the structures tested here for creep and eigenstress relaxation. As explained in the main manuscript, A and B indicate two statistically equivalent initial configurations, and $\delta$ is the fraction of particles that are inflated to generate an intense field of eigenstress.}\label{FigB} 
\end{figure}

The shear stress-strain curves for all the constructed model structures are shown in \figname\ref{FigB}. The same tests on similar C--S--H models from the literature\cite{masoero2014softmatter} provided yield stress between 120 and 300 MPa and yield strain between 0.04 and 0.06, depending on the packing density of the system. Similar values are obtained here and are plotted in \figname\ref{FigB}.

The curves in \figname\ref{FigB} support our choice of performing creep simulations using: (i) an applied stress $\tau_{xy} =40$ MPa, which is approximately one order of magnitude smaller than the yield stress - this means that all structures are far from the plastic regime; (ii) shear strain oscillations $\Delta \gamma_{xy}= \pm 0.03$ that, added to the strain corresponding to the applied $\tau_{xy}$, bring the structure close to but not past yielding. The next section will show that strain oscillations nearing the yield point provide indeed the most effective relaxation of eigenstress without causing system-wide damage.

\section{Regimes of relaxation and damage} \label{AppC}

\begin{figure}[h]
\includegraphics[width=0.7\columnwidth]{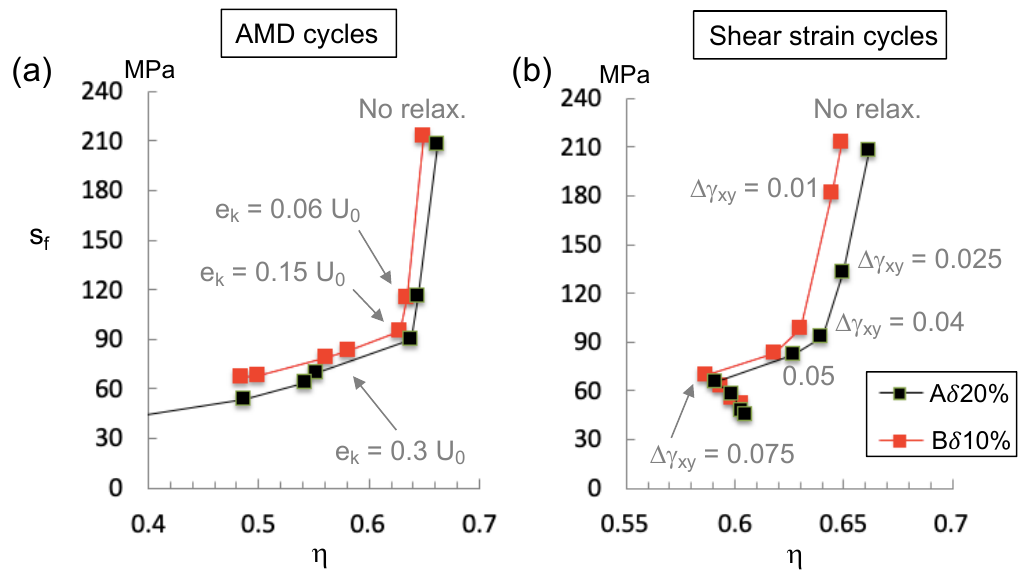}
\caption{Effect of perturbation intensity on two model structures and for the two protocols used in the main manuscript: (a) accelerated molecular dynamics, AMD, with random agitations of intensity $e_k$ (average kinetic energy per particle), and (b) cycles of shear strain oscillations of magnitude $\Delta\gamma_{xy}$. Relationships are drawn between final average tensile eigenstress $s_f$ and final packing density $\eta$, for various intensities of $e_k$ and $\Delta\gamma_{xy}$.}\label{FigC} 
\end{figure}

As presented in \secname\ref{SecMethCreep}, the undriven ($\tau_{xy} = 0$) relaxation of eigenstress has been carried out using specific values of random agitation or shear strain oscillations. Those values have been chosen after exploring various possible intensities of agitations and oscillations, to understand how these affected the final values of the eigenstress $s_f$ and the final packing density $\eta$ of the relaxed structures. \figname\ref{FigC} shows the results of such explorations for two sample structures: one of type A with a fraction $\delta = 20\%$ of inflated particles, and one of type B with $\delta = 10\%$. Similar results emerge also from all the other configurations in the manuscript.

\figname\ref{FigC} shows that there are two regimes of relaxation. When perturbations are weak, \viz small $e_k$ or $\Delta\gamma_{xy}$ depending on the protocol, the final eigenstress $s_f$ decreases significantly while the packing density $\eta$ does not change much. This is a desirable regime of relaxation, in which the structural alterations are limited to small local rearrangements. Vice versa, intense perturbations cause large changes in packing density with an evident change of regimes in the $s_f(\eta)$ plots in \figname\ref{FigC}. The perturbations marking the transitions from one regime to the other, with intensity $e_k=0.15 U_0$ in \figname\ref{FigC}.a and $\Delta\gamma_{xy}=0.04$ in \figname\ref{FigC}.b, maximise stress relaxation without inducing extensive damage (lowest $s_f$ while keeping small changes of packing density $\eta$). Therefore, these perturbations have been used in main manuscript to simulate undriven relaxation.

\end{document}